# WEST operation with real time feed back control based on wall component temperature toward machine protection in a steady state tungsten environment.

R. Mitteau, C. Belafdil, C. Balorin, X. Courtois, V. Moncada, R. Nouailletas, B. Santraine, and the WEST team

CEA, IRFM, F-13108 Saint-Paul-Lez-Durance, France
West team is defined in http://west.cea.fr/WESTteam

## Abstract

A real time Wall Monitoring System (WMS) is used on the WEST tokamak during the C4 experimental campaign. The WMS uses the wall surface temperatures from 6 fields of view of the Infrared viewing system. It extracts the raw digital data from selected areas, converts it to temperatures using the calibration and write it on the shared memory network being used by the Plasma Control System (PCS). The PCS feeds back to actuators, namely the injected power from 5 antennae's of the lower hybrid and ion cyclotron resonance radiofrequency (RF) heating systems. WMS activates feed back control 63 times during C4, which is 14% of the plasma discharges. It activates mainly as the result of a direct RF loss to the upper divertor pipes. The feedback control maintains the wall temperature within the operation envelope during 97% of the occurrences, while enabling plasma discharge continuation. The false positive rate establishes at 0.2%. WMS significantly facilitated the operation path to high power operation during C4, by managing the technical risks to critical wall components.

## 1. Introduction : Wall monitoring systems, and WEST mission

Wall Monitoring Systems (WMS) are developed in magnetic fusion machines, toward securing plasma operation through wall damage avoidance [1-4]. ITER has already integrated the need for active wall safety based on Infrared (IR) imaging diagnostic [5]. WMS has become a key asset for fusion machine operation, as internal Plasma Facing Components (PFCs) have evolved to high tech actively cooled metallic walls. Tokamak (and stellarator) walls are the thermal power sink of the fusion reactor. Current medium size machines and the next step ITER transfer 10-140 MW to the cooling system through the wall. Today's wall PFCs heat up to stable temperatures of about 1000-2000 K during plasma operation. They are among the hottest high-tech industrial components operating in steady state, along with gas turbine blades and industrial furnaces. Thermal management incidents or unstable plasma events may cause abnormal hot spots at the wall, which can evolve to a component damage, with potential consequences on machine availability. Incidents related to thermal management include local tile melting, loss of armour tiles, up to in-vessel water leak for the most severe accidents.

Historically, wall monitoring in magnetic fusion machines is achieved thanks to human expertise, by monitoring the hot spots observed on the IR films after plasma discharge. The IR movies are analysed by a PFC Protection Officer (PPO) operating in the control room. The PPO is part of the operation team. He intervenes on the plasma discharge programming, in case maximum component temperatures are approached or exceeded, or if abnormal thermal events do occur on the previous discharge. This intervention happens during the pre-pulse dialog with the session leader and the scientific coordinator. Human analysis has the major drawback of happening after the plasma discharge, a characteristic that becomes increasingly limiting as the fusion devices shift to long pulse steady state operation. Steady state plasma operation in an actively cooled metallic environment benefits from the ability to process wall hot spots in real time, requiring automatized monitoring systems. The human analysis safety branch is also impaired by the challenge of having to analyse the large amount of data produced by multiple viewing diagnostics during longer plasma discharges, giving another justification to the progressive shift to an automatized WMS.

The WMS involves many processes, including both active and passive processes. Active process rely on real-time processing of diagnostic data, while passive processes involve comparing pulse programmed parameters to pre-existing knowledge, typically defined as an operation envelope. Automatized & active wall monitoring has the ability to manage unforeseeable thermal incidents such as abnormal heat losses or deteriorating components occurring unexpectedly during the plasma discharge. It is hence effective at achieving wall protection, especially for thermal events which timescale is incompatible with human operators (of the order of seconds).

Among all possible wall temperature measurement methods, the infrared imaging system is promising for automatized & active wall monitoring. It is able to produce data on a large surface area, and directly at the component surface (without the time lag that would result from an embedded sensor). Infrared imaging systems rely on sophisticated optical endoscopes and cameras [6 - 8]. An automatized & active wall safety using analog IR cameras was already pioneered at Tore Supra in the early 2000' [6]. Tore Supra has now become WEST, a tungsten divertor steady state medium size tokamak preparing ITER operation. WEST operates ITER-grade divertor components (Figure 1), and investigates long pulse H-modes in an essentially tungsten environment [9].

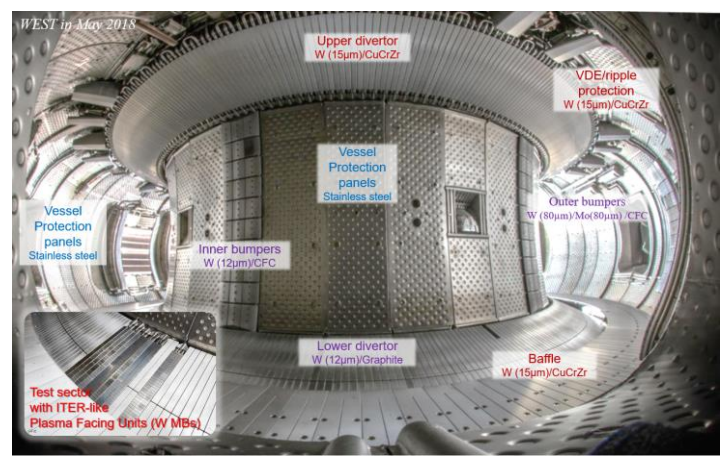

*Figure 1 : West wall and plasma facing components*

The WEST 2019 C4 campaign aimed at progressing toward the H-mode, which necessitated extending the operation domain both in power and duration. WEST is solely powered by RF heating systems : two lower hybrid launchers LH1 and LH2 (design 3 + 4 MW steady state) and three Ion cyclotron resonance heating antennae's IC1, IC2 and IC3 (design 3MW steady state, 9 MW-30s). The former Tore Supra experience is that these RF heating systems are associated to possible direct power losses to the wall. While these wall losses are a small fraction of the injected power, typically less than 10%, they can occasionally be localised spatially, hence carry a significant power load density relative to the component allowance. There is good knowledge on the underlying plasma mechanisms to these losses, however precise localisation and occurrence of these losses aren't predicted with enough confidence for ensuring hot-spot free operation. The need for an operating WMS is exacerbated in this context. An effective WMS is finalised and put in service during the S3 shutdown in early 2019 before the C4 experimental campaign. It is commissioned and qualified for operation during the initial low power operation phase of C4. Section 2 provides a description of the WEST WMS system, along with the features of the IR viewing system being relevant to WMS. The section 3 describes the main WMS activation scheme during C4, and section 4 lists statistical elements of WMS activations over C4.

## 2. WEST WMS and Infrared diagnostic system description

WEST WMS processes the wall temperatures obtained from the infrared imaging diagnostic. From the start of the WEST project, wall temperature monitoring is identified as a key project domain, because of the combined challenges of operating a tungsten wall with active cooling. An extensive Infrared system of 12 viewing lines covering about half of the WEST internal wall is built and operated [10]. Each IR camera delivers 640 x 512 x 16 bits images at 50Hz. Sensor wavelengths, filters selection, exposure times are detailed in [10]. A vertical and a horizontal channels are shown in Figure 2. There is also a wide angle tangential IR field of view, with the ability to provide a large view of the vessel interior (shown later in this paper, Figure 7).

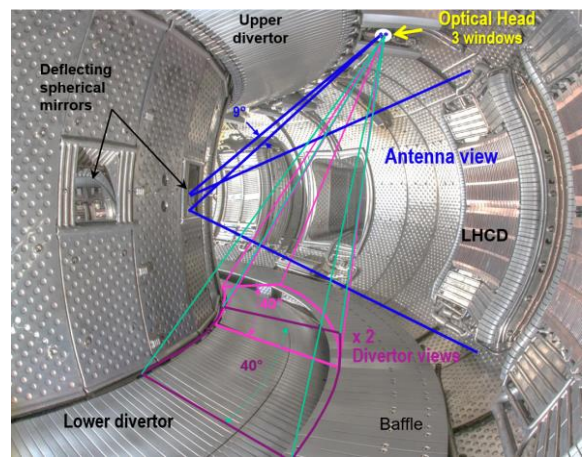

*Figure 2 : West vertical and horizontal fields of views of the Infrared endoscopes, superimposed onto a tangential view of the internal chamber of WEST fitted with internal components*

The IR diagnostic encompasses the cameras, the optical endoscopes, and possibly windows and mirror [10]. It is calibrated in the infrared laboratory using blackbody references before the experimental campaign. The calibration scope includes both calibration of the infrared cameras themselves, as well as the calibration of the optical transmissions of the endoscopes and possible windows and mirrors. The management of the emissivity of the observed surfaces in the field of view is done elsewhere : the infrared emissivity of observed in-vessel components is a scientific topic in itself, as the emissivity changes both during the campaign and depending on the observed location as the result of plasma wall interaction processes. The reader is referred to [11, 12] for quantifying these changes and a discussion on how to evaluate emissivity changes during the experimental campaign. For machine protection purposes, it is chosen to work with apparent component temperatures (that is, uncorrected for the real emissivity), and compare those measured apparent temperatures to allowable temperatures also expressed as apparent temperatures. The components allowable apparent temperatures are set assuming prudent emissivity assumptions : emissivity in the lower range given by literature survey, laboratory testing and expertise feedback. These allowables are listed (call name Tmax) in table 1 below. Hence the real time conversion from digital levels to surface temperatures is done in real time with the assumption of back body surfaces.

WEST WMS in 2019 relies on Region of Interest (ROIs). Theses ROIs are pre-defined areas, fitted to components. WMS is able to process up to 8 ROIs per field of view and is built for up to 12 fields of view. During C4, 6 fields of view are activated on WMS : the wide angle tangential view, and the five antenna views. The wide-angle view ROIs include the lower divertor and baffle, the upper divertor, the upper divertor pipes (code WA/Udiv pipes, Figure 7 and Table 1), the ripple losses protection plate and the inner bumper. All viewed area which isn't included in a pre-existing ROI is actually part of a so-called "Security ROI". The security ROI isn't used for feedback control. It is used during post-discharge analysis as a sanity check to confirm that no abnormal hot spot happened on secondary wall components. The WA/Udiv ROI is the only ROI used for feed back control on the tangential wide angle field of view.

Several ROIs are defined on each antenna view. All antennaes have left and right lateral side limiters, analog to outer poloidal bumpers, serving the role of protecting the antenna from the heat power circulating in the scrape-off layer. They are referred as "antenna_name-side limiters" (Figure 3 and Table 1). The RF power emitting area in the centre of the antenna (LH grill or faraday screen) are also monitored with dedicated ROIs, according to their allowables (the temperature allowables depend on material and cooling arrangement, per each sub-area). All these ROIs are used for feed-back control. A security ROI containing observed the zone of the field of view which is not included in a pre-existing ROI is also defined, but it is not used for feed-back control.

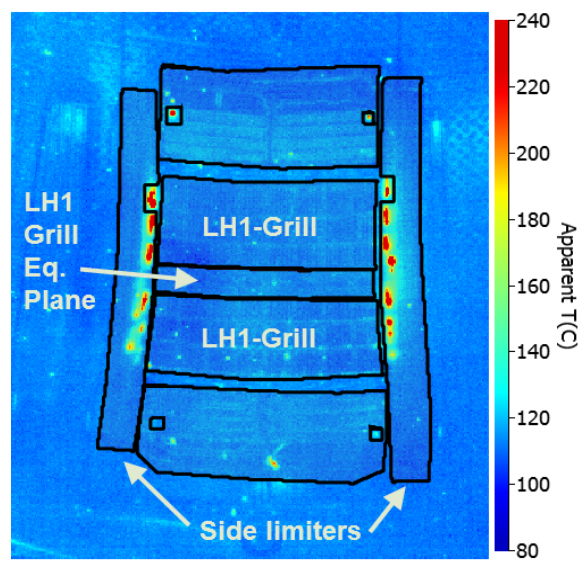

*Figure 3 : Illustration of 3 ROIs on an antenna scene - the raw image was edited for suppressing some plot artefacts, while not changing the meaningful information.*

The temperature extraction from the ROI and the conversion to apparent temperature is done in a dedicated FPGA board in the acquisition units. These boards continuously send temperature images to the WMS through optical fibbers based on the UDP protocol.

The WMS itself is essentially a fast computer (Figure 4), connected to the PCS through real-time shared memory network. The WMS unit is a 24 cores computer for parallel processing of multiple channels. Each CPU manages the acquisition and real-time processing of a unique diagnostic or IR video, while one CPU is responsible for communication with the pulse-sequencing unit, part of WEST acquisition framework. While the WMS currently processes only IR data streams, it has the capability of crossing IR data with other diagnostics such as impurity monitoring, water calorimetry and magnetic measurements for advanced management of wall safety issues.

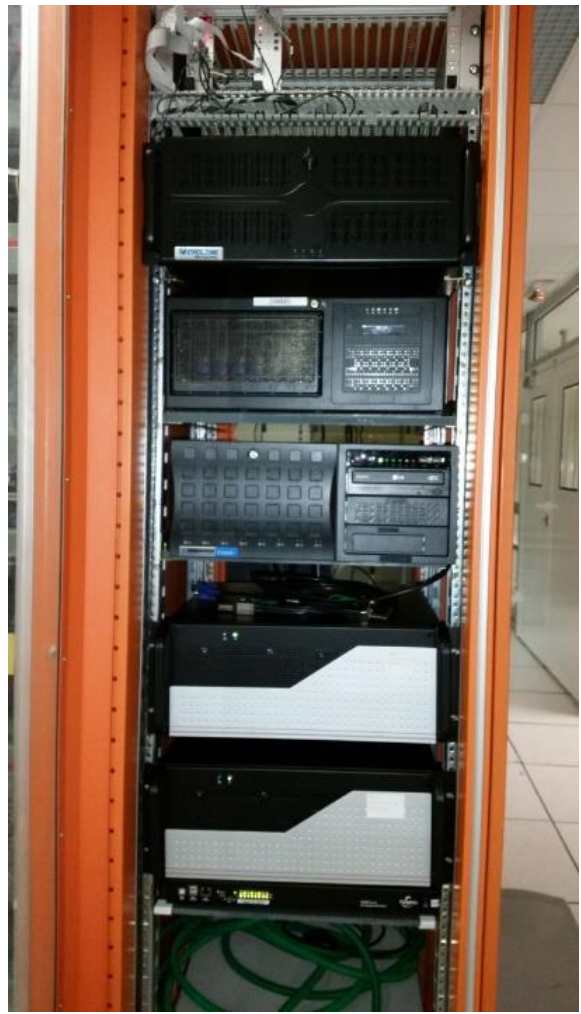

*Figure 4 : WMS electronic rack*

The ROIs maximum temperatures are written on the real-time shared memory network, and made available to the PCS for feedback control.

| ROI | $T_{alarm}$ | $T_{max}$ | LH1 | LH2 | IC1 | IC2 | IC3 |
|---|---|---|---|---|---|---|---|
| LH1-Grill | 220 | 270 | X | | | | |
| LH1-Grill eq. Plane | 550 | 600 | X | | | | |
| LH2-Grill | 220 | 270 | | X | | | |
| LH2-Grill eq. Plane | 550 | 600 | | X | | | |
| WA-Udiv Pipes | 235 | 275 | X | X | | | |
| LH1-Side Limiter | 650 | 700 | X | X | X | X | X |
| LH2-Side Limiter | 650 | 700 | X | X | X | X | X |
| IC1-Side Limiter | 650 | 700 | X | X | X | X | X |
| IC2-Side Limiter | 650 | 700 | X | X | X | X | X |
| IC3-Side Limiter | 650 | 700 | X | X | X | X | X |

*Table 1 : ROI to actuators connection table and associated temperature thresholds*

In 2019, the actuators are set to be the heating power of the RF heating systems (lower hybrid and Ion cyclotron resonance heating systems). Other actuators such as gas feeding, or magnetic control are technically possible but haven't been investigated so far.

The control law is identical for all channels. It is illustrated graphically in Figure 5 for the upper divertor pipes, which power flux density depends essentially on the heat power injected by the lower hybrid heating system. $T_{alarm}$ is set to 235°C and $T_{max}$ to 275°C. These temperatures are smaller compared to component customary allowables, but one shall remember that these are actually apparent temperatures, calculated with a prudent emissivity of $\varepsilon = 0.3$ and a correction factor due to the small size of the heating area compared to the spatial resolution. Below $T_{alarm}$, the LH power is injected according to the prescribed pre-discharge power profile. Between $T_{alarm}$ and $T_{Max}$, a linear trimming law is implemented : the injected power is reduced by a proportional low between $T_{alarm}$ and $T_{max}$. For example, if the pipe temperature sets à 15% inside the interval [$T_{alarm}$-$T_{max}$], then the LH power is reduced by 15%. Above $T_{max}$, the reduction is 100%, meaning that LH power injection is cancelled.

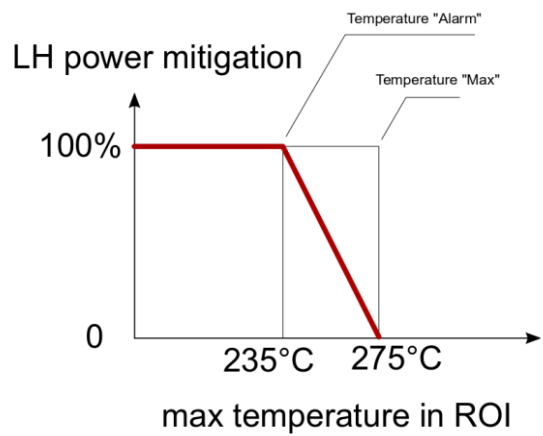

*Figure 5 : Control law, illustrated for the action of upper pipe ROI on the LH system*

The actual temperature thresholds are defined per each ROI, given in Table 1. They are set according to WEST operating instructions, themselves depending on the component design. Table 1 also defines the connection matrix : which ROI acts on which actuator. Typically, a given antenna emitting area feeds back on its own power, as illustrated for LH grills in Table 1. The WA/Udiv pipes are heated by the combined power of the whole LH system, hence the control matrix reduces both powers of LH1 and LH2 antennae's. The side limiters are heated by the total conductive power of the scrape off layer. Hence side limiter ROIs reduces the injected power of all heating systems.

For being exhaustive, it shall be mentioned that WEST IR WMS works actually also with a hard wired central interlock system. This operation scheme is relevant to that planned for ITER (Figure 6, reproduced from [13]) : the first safety threshold triggers control mechanisms by the PCS, attempting at sustaining the plasma within pulse control allowable. When control is unsuccessful and a second threshold is overshot, the interlock triggers an exception handling workflow.

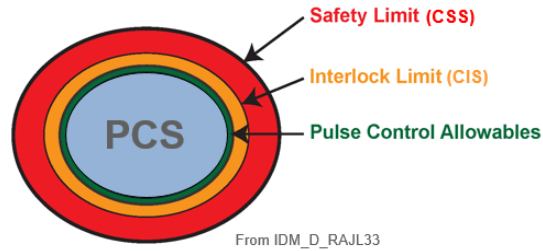

Figure 6 : Nested control envelopes, CIS stands for Central Interlock System, CSS for Central Safety System

WEST IR hard wired central interlock was monitored digitally during C4, but wasn't interfaced to the plasma fast shutdown trigger.

## 3. Detailed description of an activation occurrence

The IR based feedback control activated a number of times during the C4 campaign in 2019 (the full statistics is the object of section 4). The most frequent overpassing of $T_{alarm}$ happens on the ROI located on the upper divertor pipe WA/ Udiv pipes (Figure 7). This section 3 describes the activation sequence.

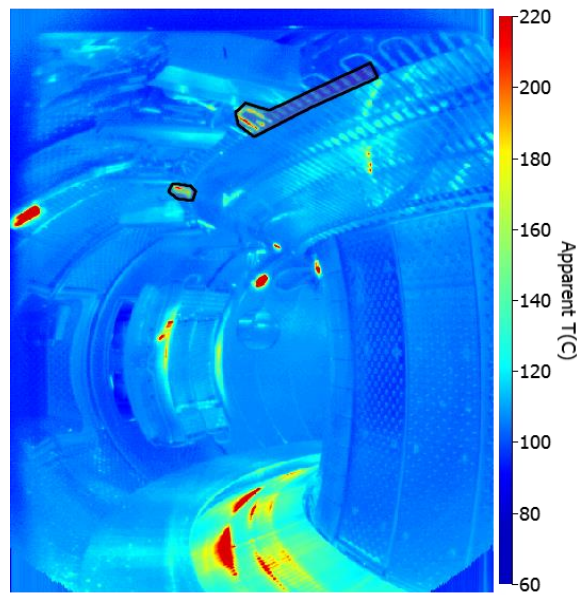

Figure 7 : WEST wide angle view. The black lined contour at the top is the upper divertor pipe ROI.

The upper divertor pipe is heated by suprathermal electron losses, occurring during lower hybrid heating, for certain plasma conditions - typically at lower plasma density. This power loss channel is well characterised [14]. It is not further described here, the reader is referred to [14] for a complete description and discussion on this power loss channel. At WEST, this heat load is absorbed by a dedicated dump plate, but the load tail may reach an actively cooled steel pipe of the upper divertor. Occasionally, the heat load density on the pipe exceeds the pipe allowable. The mechanical load resulting from the thermal gradient that sets in the pipe wall is controlled by the surface temperature. The maximum apparent temperature $T_{max}$ is set to 275°C, corresponding to a real temperature of 465°C (the prudent assumption is of a steel emissivity of 0.3). This allowable is set to keep the steel strain range within the acceptable usage factor. An excessive accumulation of strain cycles would exceed the budget lifetime, and could ultimately cause a pipe crack, hence a possible in vessel water leak.

An especially illustrative activation occurred during discharge #55828, a high power (4 MW) long pulse (20s) discharge where 3 feed-back control occurrences activated during the same power plateau. The Figure 8 illustrates these activations.

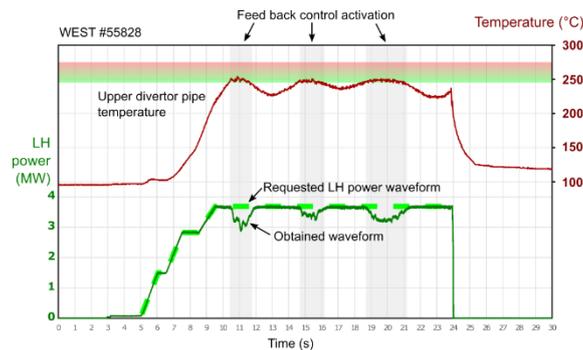

*Figure 8 : LH power mitigated by WMS during #55828*

The dashed green curve is the pre-programmed LH power profile. The power scale is on the left axis, in green too, up to 4 MW. It is ramped up from 5 to 9.5 seconds, then a plateau at 3.8 MW is requested. The pipe apparent temperature is plotted (red curve) in the upper figure. The temperature scale is on the right hand side axis, in red. The WMS control band is displayed by the green to red graded band, between 235 and 275°C. The pipe temperature enters the control band at 10.5, 14.7 and 18.8 seconds. The actual injected power is plotted with the continuous green curve in the plot. The LH power is reduced according to the control law. During the first activation, the LH power reduces by up to 20%. The reduction of the LH power causes an over-proportional reduction of the electron losses to the pipe. The pipe load reduces, the pipe temperature too, thanks to both temperature homogenisation within the pipe and possibly the active cooling too. As the pipe temperature passes below $T_{alarm}$, the feedback control loop allows to recover the requested LH power. The plasma pulse proceeds while staying within pipe design allowable, owing to the successful WMS control. Two further activations happen during the same discharge. In this case, the active feedback control is key to both enabling continuation of the plasma discharge, while safeguarding critical wall components.

## 4. Full statistical report and discussion on WMS feedback control

A systematic survey of WMS feed back control activations is recorded, starting from the discharge #55450 (3-October-2019) where the WMS feed back control is qualified for routine operation. The C4 experimental campaign runs until #55987 (15-November-2019). Over this period, 443 significant discharges are achieved, most of them being L-mode discharges (significant discharges exclude technical tests and discharges lasting less than 2 seconds). WMS feedback control activates on 60 discharges, representing 14% of the significant discharges. The distribution of activations according to the ROIs is given in Table 2.

|  | WA-Udiv pipes | LH2 grill | LH1 grill | Antennae's Side limiters |
|---|---|---|---|---|
| Num. of activations | 51 | 9 | 3 | 0 |
| Percentage | 81% | 14% | 5% | 0% |

*Table 2 : Statistics of WMS ROI activations*

The total number of activation is 63. This number is larger than the total number of 60 plasma discharges, because some discharges experience feedback control activations by more than one ROI. Multiple activations by the same ROI within the same plasma discharge, such as the one described in section 3, are counted as a single activation. During 61 activations out of the total 63, WMS succeeds at maintaining the component below $T_{max}$ by trimming the injected power, while keeping the plasma running. The success rate establishes at 97%.

There are two occurrences where the component surface temperature exceed the $2^{nd}$ threshold $T_{max}$. The first occurrence of exceeding $T_{max}$ happens on a flying dust passing in the camera field of view (#55749). This dust is a chip of the tungsten coating teared off the side limiter. Normally, a hot dust shouldn't trigger power reduction. Such an event is called 'false positive' in detection vocabulary. One occurrence of false positive over 443 discharges corresponds to a false positive rate of 0.2%, an excellent result by control standards. The false positive rate is expected to increase when more ROIs are activated, particularly on the actively cooled divertor being installed for the C5 experimental campaign. Future work aims at using computer vision techniques toward being able to manage false positives. The work plan includes building data pipelines with computer vision techniques involving pattern recognition processes, event detection and classification, toward an artificial intelligence management of the hot spots and thermal events. Key goals are to be able to detect flying objects such as loose flakes and dusts, as well as photonic reflections. Another goal is to be able to catch 'true negatives', namely thermal events being relevant for wall safety, happening below alarm temperature thresholds.

The second occurrence exceeding $T_{max}$ is caused by a fast heating event on the WA-Udiv pipes (#55822). The feedback control activates but the heating rate is faster than the LH power reduction rate. $T_{max}$ is exceeded after 60 ms, that is 3 video cycles after $T_{alarm}$. A hardwired interlock trigger is recorded, but the plasma discharge proceeds as the trigger is not connected to plasma emergency shutdown. The future work plan is to better characterise the dynamics of the heating and of the feed-back control, and possibly connect the interlock to the emergency plasma shutdown. This event is also a demonstration of the soundness of the nested envelope control principle.

## 5. Conclusion and future prospects for the WMS

A Wall Monitoring System (WMS) is activated on the WEST tokamak during the C4 experimental campaign. The WMS uses real time data from 6 fields of view of the Infrared diagnostic, processes the data and write the relevant wall temperatures on the shared memory network being used by the plasma control system (PCS). WEST PCS uses the wall temperatures, and feedbacks control to actuators, namely the injected power of the 5 radiofrequency (RF) heating systems. The WMS activates 63 times during C4, mainly as the result of a direct RF heated particle loss to the upper divertor pipes. The feedback control maintains the pipe temperature within the operation envelope during 97% of the occurrences, while permitting continuous plasma operation. Lessons learned include the need for further developing the control system by activating more ROIs and actuators, and improving the knowledge of the dynamics of fast events. It also demonstrates the interlock usefulness even with active control loops, as well as the need to adding intelligence and expertise to the automated feedback control loops though imaging techniques for avoiding false-positives.

# 7. List of figures and tables